\documentclass[aps,prd,superscriptaddress,showpacs,twocolumn,
nofootinbib,eqsecnum]{revtex4}

\usepackage{amsmath,amsfonts,amssymb}
\usepackage{graphics}
\usepackage{dcolumn}

\allowdisplaybreaks
\begin{document}

\title{A skeleton approximate solution of the Einstein field equations\\
for multiple black-hole systems}

\author{Guillaume Faye}
\email{G.Faye@tpi.uni-jena.de}
\affiliation{Theoretisch-Physikalisches Institut,
Friedrich-Schiller-Universit{\"a}t, Max-Wien-Platz 1, 07743 Jena, Germany}

\author{Piotr Jaranowski}
\email{pio@alpha.uwb.edu.pl}
\affiliation{Institute of Theoretical Physics, University of
Bia{\l}ystok, Lipowa 41, 15-424 Bia{\l}ystok, Poland}

\author{Gerhard Sch{\"a}fer}
\email{G.Schaefer@tpi.uni-jena.de}
\affiliation{Theoretisch-Physikalisches Institut,
Friedrich-Schiller-Universit{\"a}t, Max-Wien-Platz 1, 07743 Jena, Germany}

\date{\today}


\begin{abstract}
  
  An approximate analytical and non-linear solution of the Einstein field
  equations is derived for a system of multiple non-rotating black holes.  The
  associated space-time has the same asymptotic structure as the
  Brill-Lindquist initial data solution for multiple black holes.  The system
  admits an Arnowitt-Deser-Misner (ADM) Hamiltonian that can particularly
  evolve the Brill-Lindquist solution over finite time intervals.  The
  gravitational field of this model may properly be referred to as a {\em
    skeleton} approximate solution of the Einstein field equations.  The
  approximation is based on a conformally flat truncation, which excludes
  gravitational radiation, as well as a removal of some additional
  gravitational field energy. After these two simplifications, only source
  terms proportional to Dirac delta distributions remain in the constraint
  equations. The skeleton Hamiltonian is exact in the test-body limit, it
  leads to the Einsteinian dynamics up to the first post-Newtonian
  approximation, and in the time-symmetric limit it gives the energy of the
  Brill-Lindquist solution exactly.  The skeleton model for binary systems may
  be regarded as a kind of analytical counterpart to the numerical treatment
  of orbiting Misner-Lindquist binary black holes proposed by Gourgoulhon,
  Grandcl{\'e}ment, and Bonazzola, even if they actually treat the corotating
  case.  Along circular orbits, the two-black-hole skeleton solution is
  quasi-stationary and it fulfills the important property of equality of Komar
  and ADM masses.  Explicit calculations for the determination of the last
  stable circular orbit of the binary system are performed up to the tenth
  post-Newtonian order within the skeleton model.

\end{abstract}

\maketitle

\section{Introduction}

The description of the motion of binary black holes within the Einsteinian
theory of gravity is quite a challenging problem and is likely to be finally
solvable by numerical means only.  However, it seems that the full success
will occur in a far future.  It has turned out in the past that analytical
developments did always influence numerical investigations, hence further
progress on the analytical side is certainly desirable.  In this paper we
shall present an analytical solution of truncated Einstein equations for
systems of non-spinning point-like objects, which shows many aspects of past
numerical models.

Most of the numerical computations of quasi-stationary initial data for
black-hole binaries are based on the assumption that the space metric is
conformally flat \cite{SU01,GGB01a,GGB01b}. They often assume as well the
existence of an approximate helical Killing vector which permits to elaborate
some manageable formulations (also see \cite{C03}). Recently, the
time-symmetric initial data of Brill-Lindquist \cite{BL63} were generalized by
adding a non-conformally flat contribution that incorporates pieces of
information provided by perturbative post-Newtonian calculations \cite{B03}.
The present paper does not follow this line, but rather keeps the spirit of
former numerical simulations.  We essentially perform two simplifications: on
the one hand, we adopt the often-used assumption of conformally flat space
metric; on the other hand, we attribute the non-linear gravitational field
energy to point-like sources, which are modeled as Dirac delta distributions.
We solve the resulting truncated Einstein equations analytically and get what
we call, in reference to the point-like character of the source support, the
{\em skeleton} solution.  The calculations are carried out for $\mathcal{N}$
bodies in a mathematically sound way by working from the beginning in
$(d+1)$-dimensional space-time and taking the limit $d\to3$ in the final
expressions, so that the Dirac delta distributions modeling the sources are
consistently handled \cite{DJS01b}.

Our space metric and extrinsic curvature may be regarded as some
generalization of the multiple black hole Brill-Lindquist initial value
solution \cite{BL63} to arbitrary black-hole momenta preserving its original
asymptotic structure. Moreover, the energy function can be used as generator
of the dynamical evolution of the system.  We shall refer to it as the
skeleton Hamiltonian associated to the full Arnowitt-Deser-Misner (ADM)
Hamiltonian.  It agrees with the exact Hamiltonian \emph{(i)} in the test body
limit, \emph{(ii)} at the first post-Newtonian (1PN) approximation, i.e., at
the first order in powers of $1/c^2$ ($c$ denoting the speed of light) beyond
the Newtonian dynamics, and \emph{(iii)} in the limit of special relativity in
absence of gravity.  The skeleton gravitational field is given explicitly, as
well as the equations of motion.

As a first application, we shall investigate the location of the last stable
(circular) orbit (LSO) for binary systems.  In order to reach a proper
accuracy, the last-stable-orbit parameters are worked out up to the tenth
post-Newtonian (10PN) approximation within the skeleton model.  The comparison
of second and third post-Newtonian results derived from the skeleton model
with the corresponding quantities computed in the full Einstein theory
delivers a quantitative measure for the missing terms in the skeleton
approach.

\section{Description of the black-hole dynamics in ADM formalism}

We are interested in a system of $\mathcal{N}$ black holes interacting
gravitationally.  Its description will be achieved within the ADM Hamiltonian
framework using a $(d+1)$ splitting of space-time.  The spatial indices are
denoted by Latin letters and vary from $1$ to $d$, whereas the space-time
indices are Greek and vary from $0$ to $d$.  In a given chart,
$(\mathbf{x}=x^i,\,t=x^0/c)$, the motion of the $A$th black hole is described
by a curve $x^i_A(t)$, for $A=1,\,\ldots,\,\mathcal{N}$.  The conjugate
momentum of the position $x^i_A$ is called $p_{Ai}$.  The constant mass
parameter of the $A$th black hole is denoted by $m_A$.  The Hamiltonian
variables associated to the gravitational field are taken to be (i) the space
metric $\gamma_{ij}$ induced by the full space-time metric $g_{\mu\nu}$ on the
hypersurface $t=\text{const}$, and (ii) its conjugate momentum $(c^3/16\pi
G)\pi^{ij}$ where $G$ is the Newtonian gravitational constant.  The other
components of $g_{\mu\nu}$ are parametrized by the lapse and shift
functions: $N\equiv(g_{0i}g_{0j}\gamma^{ij}-g_{00})^\frac{1}{2}$ and
$N_i\equiv g_{0i}$, with $\gamma^{ij}$ standing for the inverse of
$\gamma_{ij}$.  It is also useful to introduce the shift vector
$N^i\equiv\gamma^{ij} N_j$.  The spatial indices will be raised and lowered
with the help of $\gamma^{ij}$ and $\gamma_{ij}$, respectively.

We now assume that the ``center-of-field'' $\mathbf{x}_A(t)$ of object $A$ can
be modeled in the fictitious space-time associated with the flat metric by
means of the Dirac distribution
$\delta_A\equiv\delta(\mathbf{x}-\mathbf{x}_A)$.  This prescription is known
to lead to the Brill-Lindquist Hamiltonian when the black hole velocities are
set to zero, provided the mass parameters $m_A$, $A=1,\,\ldots,\,\mathcal{N}$,
are identified with the Brill-Lindquist masses \cite{JS00b, JS02}.  It also
leads to the correct 2PN dynamics \cite{D83, IFA01} for compact binaries,
which can itself be recovered by means of the extended-body approach
\cite{K85}.  Moreover, the singularities generated by the distributions
$\delta_A$ have shown to be curable by dimensional regularization at the 3PN
order \cite{DJS01b}.  Similarly, we shall see in the present work that, for an
appropriate choice of the space dimension $d$, the $\delta_A$-type sources do
not generate any ambiguity, so that the 3-dimensional quantities can be
obtained by dimensional regularization.  Finally, this point-particle approach
allows to bypass delicate investigations about the behavior of the
gravitational field near the objects.

In an asymptotically flat space-time of dimension $d+1$, the Hamiltonian of
$\mathcal{N}$ point masses provided by the ADM formalism reads
\cite{ADM62,RT74,H85}
\begin{multline}
\label{eq:full_Hamiltonian}
H = \frac{c^4}{16\pi G} \Bigg( \int d^d\mathbf{x}
\left(N \mathcal{H} + N^i \mathcal{J}_i\right)
\\
+ \int d^d\mathbf{x}  \,\, \partial_i 
(\partial_j\gamma_{ij}-\partial_i\gamma_{jj}) \Bigg),
\end{multline}
where $\partial_i\equiv\partial/\partial x^i$. The super-Hamiltonian density
$\mathcal{H}$ appearing under the integral sign is defined by
\begin{subequations}
\begin{align}
\label{eq:super_Hamiltonian}
\mathcal{H} \equiv & -\text{R}\sqrt{\gamma} + \frac{1}{\sqrt{\gamma}}
\left( \pi^i_{\,j}\pi^j_{\,i} - \frac{1}{d-1} (\pi^i_i)^2 \right)
\nonumber\\
& + \frac{16\pi G}{c^2} \sum_{A=1}^\mathcal{N}
\left( m_A^2 +\frac{\gamma^{ij}p_{Ai}p_{Aj}}{c^2} \right)^\frac{1}{2} \delta_A,
\end{align}
while the supermomentum density $\mathcal{J}_i$ is given by
\begin{equation}
\label{eq:supermomentum}
\mathcal{J}_i \equiv - 2 \partial_j \pi^j_{\, i} + \pi^{kl} \partial_i 
\gamma_{kl}
- \frac{16 \pi G}{c^3} \sum_{A=1}^\mathcal{N} p_{Ai} \delta_A.
\end{equation}
\end{subequations}
In Eq.\ \eqref{eq:super_Hamiltonian} $\text{R}$ denotes the space curvature of
the hypersurface $t=\text{const}$ and $\gamma$ is the determinant of the
matrix $\gamma_{ij}$. As the lapse and shift functions are Lagrange
multipliers, the variation of the Hamiltonian \eqref{eq:full_Hamiltonian} with
respect to them leads to the constraint equations
\begin{equation}
\label{eq:constraint_equations}
\mathcal{H} = 0, \quad \mathcal{J}_i = 0.
\end{equation}

Before solving Eqs.\ \eqref{eq:constraint_equations}, we must first fix the
coordinate system. It is convenient to work in the so-called ADM transverse
trace-free (ADMTT) gauge \cite{OOKH74a,S85} defined by the two conditions
\begin{equation}
\partial_j\left(\gamma_{ij}-\frac{1}{d}\delta_{ij}\gamma_{kk}\right) = 0,
\quad \pi^{ii} = 0.
\end{equation}
In this gauge, the metric decomposes into its trace $\gamma_{kk}$, which may
be parametrized as $d\,\Psi^{4/(d-2)}$ (in order to get a simple expression
for the curvature density), and a transverse trace-free part
$h^\text{TT}_{ij}$. We thus have
\begin{subequations}
\begin{gather}
\label{eq:gamma_Einstein}
\gamma_{ij} = \Psi^\frac{4}{d-2}\delta_{ij} + h_{ij}^\text{TT}, \quad
\partial_j h_{ij}^\text{TT} = 0, \quad h_{ii}^\text{TT} = 0,
\\[1ex]
\pi^{ii} = 0.
\end{gather}
\end{subequations}
The momentum $\pi^{ij}$ needs itself to be split into a transverse trace-free
contribution $\pi^{ij}_\text{TT}$, and a rest $\tilde{\pi}^{ij}$,
\begin{equation}
\pi^{ij} = \tilde{\pi}^{ij} + \pi^{ij}_\text{TT}.
\end{equation}
These two terms are uniquely defined assuming they decay as
$1/(x^ix^i)^{\frac{d-1}{2}}$ at spatial infinity and are given as a
combination of derivatives of the Poisson integral
$\Delta^{-1}\partial_j\pi^{ij}$, and the Poisson integral of the latter
quantity, namely $\Delta^{-2}\partial_j\pi^{ij}$.

We can then try to compute $\Psi$ and $\tilde{\pi}^{ij}$ from the two
constraint equations \eqref{eq:constraint_equations} supplemented by the
boundary conditions at spatial infinity resulting from the asymptotic flatness
of space-time; notably $\Psi-1{\sim}(x^kx^k)^\frac{d-2}{2}$,
$\tilde{\pi}^{ij}{\sim}1/(x^kx^k)^{\frac{d-1}{2}}$. It is actually more
convenient to solve the Hamiltonian constraint for the quantity
$\phi\equiv\frac{4(d-1)}{d-2}(\Psi-1)$ rather than $\Psi$ to get rid off the
spurious constant at spatial infinity. We thus have
\[
\Psi = 1 + \frac{d - 2}{4 (d - 1)} \phi \,.
\] 

After solving the constraint equations \eqref{eq:constraint_equations}, we
insert the ensuing expressions for $\phi$ and $\tilde{\pi}^{ij}$ into the
right-hand side of Eq.\  \eqref{eq:full_Hamiltonian}. We get a {\em reduced} 
Hamiltonian depending on $x^i_A$, $p_{Ai}$, $h_{ij}^\text{TT}$, and 
$\pi^{ij}_\text{TT}$ only:
\begin{align}
\label{eq:reduced_Hamiltonian}
H_\text{red}
& = H_\text{red}[x^i_A,p_{Ai},h_{ij}^\text{TT},\pi^{ij}_\text{TT}]
\nonumber\\
& = \frac{c^4}{16 \pi G} \int d^d\mathbf{x}~\partial_i
(\partial_j \gamma_{ij}-\partial_i \gamma_{jj})
\nonumber\\
& = -\frac{c^4}{16 \pi G} \int d^d\mathbf{ x}~\Delta \phi.
\end{align}
This Hamiltonian contains the full information about the evolution of
the matter $(x^i_A,p_{Ai})$ and field $(h_{ij}^\text{TT},\pi^{ij}_\text{TT})$ 
variables \cite{RT74,S85}.
As the change of variables 
$(\gamma_{ij},\pi^{ij})\to(h_{ij}^\text{TT},\pi^{ij}_\text{TT})$ is not 
canonical, the Poisson brackets of $h_{ij}^\text{TT}$ and $\pi^{ij}_\text{TT}$ 
take a form involving a transverse trace-free projection operator 
\cite{RT74,S85}.

\section{Skeleton Hamiltonian} \label{sec:skeleton}

The reduced Hamiltonian $H_\text{red}$ provides the equations of motion of the
$\mathcal{N}$ bodies together with the evolution equations of the
gravitational field.  As they are strongly coupled it is hopeless to solve
them analytically, but they can be solved perturbatively using, notably, the
post-Newtonian iterative scheme \cite{S85, JS98, DJS01e98}.  The resulting
dynamics is highly reliable as long as the typical speeds remain smaller than,
say, about $0.3 c$. When binary black holes reach their last stable orbits,
the reliability of the approximation breaks down since the system enters a
strong field regime.  The only way to understand precisely this crucial stage
is to perform numerical simulations in full general relativity.  Our purpose
here is to search for a simple non-perturbative model applicable to arbitrary
strong fields, eventually leaving the Einsteinian theory but always staying
close enough to it, in order to give an element of comparison with numerical
investigations.  The space metric obtained in our approach should also be
usable as initial condition.

We shall adopt here a Wilson-Mathews-type prescription \cite{WM89, W90},
setting the non-conformally flat part of $\gamma_{ij}$, namely
$h^\text{TT}_{ij}$, equal to zero.  As $h_{ij}^\text{TT}$ is of order $1/c^4$,
the difference between our $\gamma_{ij}$ and the Einsteinian space metric
shows up at the 2PN level.  For spherically symmetric space-times the
difference is zero.  In the case where the present model is used to fix
initial Cauchy data, the condition of a conformally flat space metric needs to
hold on only one hypersurface, say $\Sigma_0: t=t_0$.  Choosing
$h_{ij}^\text{TT}$ to be zero on $\Sigma_0$ is always possible, though it is
not physically realistic in general for isolated self-gravitating systems. Our
attitude in this instance will be to accept the resulting error.  In return,
we shall be able to determine the field $\gamma_{ij}$ without assuming zero
velocities of the bodies at initial time.

We thus demand here that
\begin{equation}
\label{eq:CFC}
h_{ij}^\text{TT} = 0,
\end{equation}
and the problem reduces to solve the constraints $\mathcal{H}=0$,
$\mathcal{J}_i=0$ under this condition. The space metric $\gamma_{ij}$ is then
proportional to $\delta_{ij}$, which implies that in the supermomentum
$\mathcal{J}_i$ the term $\pi^{kl} \partial_i \gamma_{kl}$ vanishes [cf.\ Eq.\ 
\eqref{eq:supermomentum}]. The field momentum $\pi^{ij}$ relates to
$\pi^i_{\,j}$ through the identities
\begin{equation}
\pi^{ij} = \gamma^{jk} \pi^i_{\, k} = \Psi^{-\frac{4}{d-2}} \pi^i_j \,.
\end{equation}
In particular, $\pi^i_{\,j}$ is symmetric and trace-free (STF) as is
$\pi^{ij}$. The determinant $\gamma$ appearing in $\mathcal{H}$ is merely
\begin{equation}
\gamma =  \Psi^{\frac{4d}{d-2}} \,.
\end{equation}
Finally, the 3-curvature density $\text{R}\sqrt{\gamma}$ takes a remarkably
simple form \cite{L44}:
\begin{equation} \label{eq:curvature_density}
\text{R}\sqrt{\gamma} = - \Psi \Delta\phi \,.
\end{equation}
Collecting these relations together with the constraint equations 
\eqref{eq:constraint_equations}, we find:
\begin{subequations}
\label{eq:constraint_system}
\begin{align}
\label{eq:constraint_energy}
\Delta\phi = & -\Psi^{-\frac{3d-2}{d-2}}
\pi^i_{\,j}\pi^j_{\,i}
\nonumber\\[1ex]
& -\frac{16\pi G}{c^2} \sum_{A=1}^\mathcal{N} m_A \delta_A \Psi^{-1}
\left( 1 + \Psi^{-\frac{4}{d-2}}
\frac{p_A^2}{m_A^2 c^2} \right)^{\frac{1}{2}} \,,
\\
\label{eq:constraint_momentum}
\partial_j\pi^j_{\,i} = & -\frac{8\pi G}{c^3} \sum_{A=1}^\mathcal{N} p_{Ai}
\delta_A \,, 
\end{align}
with $p_A^2 \equiv p_{Ai} p_{Ai}$.
\end{subequations}

In the system \eqref{eq:constraint_system} the second equation decouples from
the first one and is solved by searching for a particular solution
proportional to the symmetric trace-free part of the derivative of a certain
vector potential $V_i$; namely,
\begin{equation}
\label{eq:mixed_momentum}
\pi^{i~\text{(part)}}_{\, j} = \text{STF} (2 \partial_i V_j) \equiv
\partial_i V_j + \partial_j V_i - \frac{2}{d} \delta_{ij} \partial_k V_k \, .
\end{equation}
The equation for the divergence $\partial_k V_k$ is obtained by applying the 
operator $\partial_{ij} \equiv \partial_i\partial_j$ 
simultaneously on the two sides of \eqref{eq:mixed_momentum}, hence $2 \Delta
\partial_k V_k = d/(d -1) \partial_{ij} \pi^i_{\, j}$. This yields immediately
$\partial_j \pi^j_{\, i} = \Delta V_i + (\frac{d}{2} - 1)/(d-1) \Delta^{-1}
\partial_{ijk} \pi^j_{\, k}$, and from the constraint
\eqref{eq:constraint_momentum} it follows that:
\begin{equation}
\Delta V_i = - \frac{8 \pi G}{c^3} \sum_{A=1}^\mathcal{N} 
\left( p_{Ai} \delta_A - \frac{d-2}{2(d-1)} p_{Aj} \partial_{ij} 
\Delta^{-1} \delta_A \right) \,.
\end{equation}
We now notice that the $d$-dimensional Laplacian of the $m$th power of $r_A 
\equiv [(x^i-x^i_A)(x^i-x^i_A)]^{\frac{1}{2}}$ satisfies
\begin{equation}
\label{eq:Laplacien_rm}
\Delta r_A^m = m (m-2 + d) r_A^{m-2}
\end{equation}
in the sense of functions, and that, in the sense of distributions,
\begin{equation}
\label{eq:Laplacien_rd_2}
\Delta r_A^{2-d} = -\frac{4\pi^\frac{d}{2}}{\Gamma\left(\frac{d-2}{2}\right)} 
\delta_A \,.
\end{equation}
As a result we may choose:
\begin{equation}
\label{eq:Vi}
V_i = \frac{G}{c^3} \frac{\Gamma(\frac{d-2}{2})}{2\pi^{\frac{d-2}{2}}}
\sum_{A=1}^\mathcal{N} \left( \frac{4 p_{Ai}}{r_A^{d-2}} - 
\frac{(d-2)p_{Aj}}{(d-1)(4-d)} \partial_{ij}r_A^{4-d} \right),
\end{equation}
which is nothing but the 1PN vector potential $V^i_{(3)}$ of Ref.\ 
\cite{DJS01b} verifying $\pi^{ij} = \text{STF}(2\partial_i V^j_{(3)}) +
\mathcal{O}(1/c^5)$.

Our second simplification consists in stating that the homogeneous solution of
Eq.\ \eqref{eq:constraint_momentum},
$\pi^{i~\text{(hom)}}_{\,j}\equiv\pi^i_{\,j}-\pi^{i~\text{(part)}}_{\,j}$, is
precisely zero, which implies $\pi^{i~\text{TT}}_{\,j}=0$.  When this
prescription is imposed on one unique hypersurface, it merely represents a
particular choice of initial data.  The resulting field momentum
$\pi^i_{\,j}=\text{STF}(2\partial_iV_j)$ is identical to the rescaled
extrinsic curvature proposed by Bowen and York in their time-asymmetric
formulation of the Cauchy problem \cite{BY80} and adopted e.g.\ in the
puncture method \cite{BB97} or various calculations of the last stable
circular orbit \cite{C94,B00,PTC00}. Although it is a mathematical solution of
the constraint equations for a conformally flat metric, it is not realistic in
the sense that it does not result from an evolution process in absence of
incoming radiation.  It does not coincide either with the Wilson-Mathews
momentum $(\pi^i_{\, j})_\text{WM}$ derived from the condition
$\dot{h}_{ij}^\text{TT}=0$:
\begin{equation}
(\pi^i_{\,j})_\text{WM} = (1+\phi) \pi^{i~\text{(part)}}_{\,j}
+ (d - 1) \tilde{\pi}^{ij}_{(5)} + \mathcal{O} \left(\frac{1}{c^6} \right) \,,
\end{equation}
where the quantity $\tilde{\pi}^{ij}_{(5)}$, defined as the $1/c^5$ part of
the solution [matching the form (3.6)] of the equation $(d - 1)\partial_j
\tilde{\pi}^{ij}_{(5)} = - \partial_j (\phi \pi^{j\, \textrm{(part)}}_{\, i})$,
is explicitly given in paper \cite{JS98} for $d=3$. The difference $(\pi^i_{\,
  j})_\text{WM} - \pi^{i~ \text{ (part)}}_{\, j}$ is only of order $1/c^5$,
which corresponds to the 2PN approximation. On the Hamiltonian level, it is of
3PN order only. One might then argue that the Wilson-Mathews condition
$h_{ij}^\text{TT} = 0$, valid for all times, i.e.\ particularly
$\dot{h}_{ij}^\text{TT} = 0$, is in contradiction to the condition $\pi^{i ~
  \text{TT}}_{\, j}=0$. However, it is not the case because the momentum
constraint equation is linear in $\pi^i_{\, j}$ and decouples from any other
field equation when the space metric is conformally flat, so that the quantity
obtained by adding an arbitrary homogeneous solution to $\pi^i_{\,j}$ still
satisfies the constraint. As our $\pi^i_{\,j}$ differs from
$(\pi^i_{\,j})_\text{WM}$ by such a homogeneous solution (i.e. a
transverse-trace free object), there is no contradiction between the two
approaches.

The integration of the Hamiltonian constraint \eqref{eq:constraint_energy} is
much more involved, principally because of the appearance of $\phi$ in the
source and the presence of a non-compact support term. On the contrary, the
compact support term is harmless. It is made of a sum of Dirac distributions
$\delta_A$ that formally determine the boundary conditions near the black
holes. By subtracting from the conformal factor the $1/r_A^{d-2}$ terms
entering $\phi$, we get an object that satisfies an equation whose source is
entirely free of $\delta_A$'s (this is the essence of paper \cite{BB97}). By
contrast, we shall keep here the constraint equation
\eqref{eq:constraint_energy} under its original form in order to determine the
dependence of the coefficient of $1/r_A^{d-2}$ on the positions $x^i_A$ and
momenta $p_{Ai}$.

To solve the Hamiltonian constraint \eqref{eq:constraint_energy} iteratively,
we have to face the delicate computation of
$\Delta^{-1}\left(\Psi^\frac{2-3d}{d-2} \pi^i_{\,j}\pi^j_{\,i}\right)$.  The
expansion of the prefactor
$\Psi^\frac{2-3d}{d-2}=\left(1+\frac{d-2}{4(d-1)}\phi\right)^\frac{2-3d}{d-2}$
in powers of $\phi$ is known to give rise to poles when $d\to3$ at the level
corresponding to the 3PN approximation \cite{DJS01b}.  In full general
relativity, these poles cancel in the Hamiltonian with similar quantities
coming from $h_{ij}^\text{TT}$, but such a cancellation cannot occur in our
case.  In particular, \emph{the 3PN approximation of the Wilson-Mathews model
  does not exist} for binary black-holes.  We must therefore treat the $\pi^2$
term differently.  Let us first note that this term enters the space metric at
the 3PN order only, since $\pi^i_{\,j}\pi^j_{\,i} = \mathcal{O}(1/c^6)$, but
contributes to the 1PN Hamiltonian [because of the global factor $c^4/(16\pi
G)$ in Eq.\ \eqref{eq:reduced_Hamiltonian}].  To be consistent with the
accuracy of the former assumption $h_{ij}^\text{TT} = 0$ which amounts to
neglect some 2PN corrections in the Hamiltonian, we have to keep at least the
leading term in the post-Newtonian expansion of
$\left(1+\frac{d-2}{4(d-1)}\phi\right)^\frac{2-3d}{d-2}
\pi^i_{\,j}\pi^j_{\,i}$, modulo a possible total space derivative.  Now, the
tensor density $\pi^i_{\,j}$ is itself proportional to the symmetric
trace-free part of the space derivative $\partial_i V_j$, so that
$$
\pi^i_{\,j} \pi^j_{\,i}
= 2 \text{STF}(\pi^i_{\, j}) \text{STF}(\partial_i V_j)
= 2 \pi^i_{\, j} \partial_i V_j \,.
$$
As a consequence, the latter expression can be split into a ``skeleton''
term $-2V_j\partial_i\pi^i_{\,j}$ with compact support, and a ``flesh''
term $\partial_i(2V_j\pi^i_{\,j})$ which takes into account the field between
the point particles. The flesh term contributes to the Hamiltonian through
the quantity:
\begin{multline*}
\frac{c^4}{16\pi G} \int d^d\mathbf{x}\, \Psi^\frac{2-3d}{d-2}
\partial_i(2 V_j \pi^i_{\, j})
\\
= \frac{c^4}{32\pi G} \frac{3d-2}{d-1} \int d^d\mathbf{x}
\Bigg( \Psi^\frac{4(1-d)}{d-2} V_j \pi^i_{\,j} \partial_i\phi \Bigg).
\end{multline*}
We are allowed to regard it as negligible in our scheme and keep the skeleton
term alone in the decomposition of $\pi^i_{\,j}\pi^j_{\,i}$. Our third 
simplification consits thus in performing the substitution
\begin{multline}
\label{eq:skeleton_approximation}
\Psi^\frac{2-3d}{d-2} \pi^i_{\,j} \pi^j_{\,i}
\to -2 \Psi^\frac{2-3d}{d-2} V_j \partial_i \pi^i_{\, j}
\\[1ex]
= \frac{16 \pi G}{c^3} \Psi^\frac{2-3d}{d-2}
\sum_{A=1}^\mathcal{N} p_{Aj} V_j \delta_A \,. 
\end{multline}

After the replacement \eqref{eq:skeleton_approximation} is made, the
right-hand side of Eq.\ \eqref{eq:constraint_energy} takes the form $\sum_A
f_A(\mathbf{x}) \delta_A$, where the $f_A$'s are some unknown functions. For
$d$ belonging to an appropriate range of values, $f_A$ is regular at
$\mathbf{x}=\mathbf{x}_A$, and $\Delta\phi=\sum_A f_A(\mathbf{x}_A)\delta_A$.
We see that in our model the source of the Hamiltonian constraint reduces to
some ``skeleton'' made of a linear combination of Dirac distributions. For this
reason, all quantities computed in the present approximation will be referred
to as skeleton quantities. Defining the $\mathcal{N}$ quantities $\alpha_A$ by
the relation
$$
-\frac{16\pi G}{c^2} \alpha_A \equiv f_A(\mathbf{x}_A), \quad 
A=1,\,\ldots,\,\mathcal{N},
$$
the function $\phi$ can be written as
\begin{equation}
\label{eq:phi}
\phi = \frac{4G}{c^2} \frac{\Gamma(\frac{d-2}{2})}{\pi^\frac{d-2}{2}}
\sum_{A=1}^\mathcal{N} \frac{\alpha_A}{r_A^{d-2}} \,,
\end{equation}
and the conformal factor reads
\begin{equation}
\label{eq:psi}
\Psi = 1 +  A_d \sum_{A=1}^\mathcal{N} \frac{\alpha_A}{r_A^{d-2}} \, ,
\end{equation}
where we have posed, in order to lighten the notation,
\[
A_d \equiv \frac{G}{c^2} \frac{d-2}{d-1}
\frac{\Gamma\left(\frac{d-2}{2}\right)}{\pi^{\frac{d-2}{2}}} \,.
\]
It is also useful to introduce the (regularized) value of $\Psi$ at the
particle position $\mathbf{x}_A$:
\[
\Psi_A \equiv \Psi(\mathbf{x}=\mathbf{x}_A)
= 1 + A_d \sum_{B \neq A} \frac{\alpha_B}{r_{AB}^{d-2}},
\]
where $r_{AB}\equiv[(x^i_A-x^i_B)(x^i_A-x^i_B)]^\frac{1}{2}$.

After making the replacement \eqref{eq:skeleton_approximation} on the
right-hand side of Eq.\ \eqref{eq:constraint_energy}, inserting there
expressions \eqref{eq:phi} and \eqref{eq:psi}, we arrive at
\begin{widetext}
\begin{align}
\label{eq:delta_phi}
-\frac{c^2}{16\pi G} \Delta\phi
= \sum_{A=1}^\mathcal{N} \alpha_A \delta_A
= & \sum_{A=1}^\mathcal{N} \Bigg\{m_A
\bigg( 1 + A_d\sum_{B \neq A}\frac{\alpha_B}{r_{AB}^{d-2}} \bigg)^{-1}
\Bigg[1 + \bigg( 1 + A_d \sum_{C \neq A} \frac{\alpha_C}{r_{AC}^{d-2}}
\bigg)^{-\frac{4}{d-2}} \frac{p_A^2}{m_A^2 c^2} \Bigg]^{\frac{1}{2}}
\nonumber\\
& + \bigg(1 + A_d \sum_{B \neq A} \frac{\alpha_B}{r_{AB}^{d-2}}
\bigg)^{\frac{2-3d}{d-2}} \frac{p_{Ai} V_{Ai}}{c} \Bigg\} \delta_A,
\end{align}
with $V_{Ai}\equiv V_i(\mathbf{x}=\mathbf{x}_A)$.
Making use of Eq.\ \eqref{eq:Vi}, we see that
\begin{equation}
p_{Ai} V_{Ai} = \frac{A_d}{2c} \sum_{B \ne A}
\bigg( \frac{3d-2}{d-2} (p_{Ai}\,p_{Bi})
+ (d-2) (n_{AB}^i\,p_{Ai})(n_{AB}^j\,p_{Bj}) \bigg) r_{AB}^{2-d}.
\end{equation}
Because of the linear independence of the distributions $\delta_A$, the
Hamiltonian constraint \eqref{eq:delta_phi} is finally equivalent to a system
of $\mathcal{N}$ algebraic equations which reads
\begin{equation}
\label{eq:alpha_A}
\alpha_A = \frac{m_A}{1+ A_d \sum_{B \neq A}\frac{\alpha_B}{r^{d-2}_{AB}}}
\left[1 + \frac{p_A^2/(m_A^2 c^2)}{\left(1+ A_d \sum_{C \neq A}
\frac{\alpha_C}{r^{d-2}_{AC}} \right)^\frac{4}{d-2}} \right]^\frac{1}{2}
+ \frac{p_{Ai} V_{Ai}/c}{\left(1+ A_d \sum_{B \neq A}
\frac{\alpha_B}{r^{d-2}_{AB}} \right)^\frac{3d-2}{d-2}},
\quad A=1,\,\ldots,\,\mathcal{N}. 
\end{equation}
\end{widetext}
Solving the above system provides in principle the values of all the
$\alpha_A$'s. These quantities represent some effective masses, as suggested
by the approximate equality $\alpha_A = m_A + \mathcal{O}(1/c^2)$. They are
functions of $x^i_A - x^i_1$ (for $A \neq 1$) and $p_{Ai}$ [defined implicitly
by Eqs.\ \eqref{eq:alpha_A}] .

The skeleton Hamiltonian reads
\begin{equation}
\label{eq:skeleton_Hamiltonian}
H = - \frac{c^4}{16 \pi G} \int d^d \mathbf{x}~\Delta \phi
= c^2 \sum_{A=1}^\mathcal{N} \alpha_A \,,
\end{equation}
which shows that the ADM mass $M_\text{ADM} = H/c^2$ is just the sum of the
$\mathcal{N}$ effective masses $\alpha_A$.  The Hamiltonian
\eqref{eq:skeleton_Hamiltonian} depends only on the matter variables:
$H=H(x_A^i-x^i_1,p_{Ai})$, so that the system is automatically conservative.
The space metric is deduced from Eqs.\ \eqref{eq:gamma_Einstein},
\eqref{eq:CFC}, and \eqref{eq:psi}:
\begin{equation}
\gamma_{ij} = \left(1 + A_d \sum_{A=1}^\mathcal{N} \frac{\alpha_A}{r_A^{d-2}} 
\right)^\frac{4}{d-2} \delta_{ij}. 
\end{equation}

By construction, $H$ and $\gamma_{ij}$ coincide with their counterparts in the
Einsteinian theory up to the 1PN order.  The test-body limit is achieved by
taking the mass of one black hole, say of label 1, much larger than the masses
of its companion: $m_1 \gg m_A$, where $A=2,\,\ldots,\,\mathcal{N}$. To get
the test-body Hamiltonian exactly, one has to let formally $m_1$ go to
infinity ($m_1 \to +\infty$) in the Hamiltonian
\eqref{eq:skeleton_Hamiltonian} while keeping $m_1/r_{1A}^{d - 2}$,
$A=2,\,\ldots,\,\mathcal{N}$, constant as well as all linear momenta $p_{Ai}$.
It follows that particularly the flesh term $\partial_i(2V_j\pi^i_{\,j})$ does
not contribute to this approximation. In absence of gravity ($G \to 0$) our
Hamitlonian has the form of the exact special-relativistic Hamitlonian for the
system of ${\cal N}$ free point particles. When $p_{Ai}=0$ for all $A$, we
recover the Brill-Lindquist solution, for which Eqs.\ \eqref{eq:alpha_A}
reduce to $\alpha_A(1+A_d\sum_{B\neq A}\alpha_B/r^{d-2}_{AB})= m_A$. As a
consequence our model also generalizes the Brill-Lindquist data.

Let us observe that the particle masses $m_A$ cannot be equated to the ADM
masses computed at the spatial infinities of the $\mathcal{N}$ black hole
sheets when one or several momenta are non-zero.  When such an identification
is performed, it leads automatically to the same relation between $\alpha_A$
and $m_A$ as in the zero momentum case, whereas the correct dynamics up to the
1PN order is known to be the one that derives from the Hamiltonian
\eqref{eq:skeleton_Hamiltonian} determined by the set of Eqs.\ 
\eqref{eq:alpha_A} with $p_{Ai} \neq 0$.  The actual dependence of the
effective masses on the $m_A$'s must take the kinetic energy of the system
into account. This indicates a relative boost of the black hole sheets with
respect to each others.  The issue is complicated by the fact that the
Bowen-York data for one single black hole do not result from a boosted
Schwarzschild field \cite{GKP99}.  This point, being still rather unclear,
needs to be investigated in detail.  We do not intend to discuss it here.

\section{Skeleton gravitational field}

In the ADM formalism, the evolution of the space metric in $(d+1)$ dimensions 
is given by [cf.\ Eq.\ (4.70) in Ref.\ \cite{H85}]\footnote{We use here
and below the usual notation for symmetrization: $2 D_{(i}N_{j)} \equiv
D_i N_j+D_j N_i $.}
\begin{equation}
\label{eq:metric_evolution}
\frac{1}{c} \partial_t \gamma_{ij}
= \frac{N}{\sqrt{\gamma}} (2 \pi_{ij} - \pi^k_{\, k} \gamma_{ij})
+ 2 D_{(i} N_{j)} \,,
\end{equation}
where the operator $D_i$ denotes the space covariant derivative. Its explicit
action on the shift function $N_i$ after symmetrization is
\begin{equation}
D_{(i} N_{j)} = \gamma_{k(i} \partial_{j)} N^k
+ \frac{1}{2} N^k \partial_k \gamma_{ij} \,.
\end{equation}

For the conformally flat 3-metric
$\gamma_{ij} = \Psi^{\frac{4}{d-2}} \delta_{ij}$, we have
$\gamma_{ij} - \frac{1}{d} \gamma_{kk} \delta_{ij} = 0$, hence
\[
\partial_t \left(\gamma_{ij} - \frac{1}{d} \gamma_{kk} \delta_{ij} \right) = 0
\,.
\]
After replacing in the above equation $\partial_t\gamma_{ij}$ and
$\partial_t\gamma_{kk}$ by the expressions following from Eq.\ 
\eqref{eq:metric_evolution} and dropping all contributions proportional to
$\pi^k_{\,k}=0$, we arrive at:
\begin{equation}
\label{eq:equation_Ni}
\text{STF} (\partial_i N^j) = - \Psi^{-\frac{2d}{d-2}} N \pi^i_{\,j} \,.
\end{equation}

The evolution equation for the field momenta $\pi^{ij}$ is known in $d$
dimensions as well [cf.\ Eq.\ (4.71) in Ref.\ \cite{H85}]. It reads
\begin{widetext}
\begin{align}
\label{eq:momentum_evolution}
\frac{1}{c}\partial_t\pi^{ij} = & -\sqrt{\gamma} \Bigg[
N \left(\mathrm{R}^{ij} - \frac{1}{2} \gamma^{ij} \mathrm{R} \right)
- D^i D^j N + \gamma^{ij} D_m D^m N \Bigg]
+  \frac{N}{\sqrt{\gamma}} \Bigg[ \pi^{ij} \pi^k_{\, k}
- 2 \pi^i_{\, k}\pi^{kj} + \frac{1}{2} \gamma^{ij} \left(\pi^{kl} \pi_{kl}
- \frac{1}{2} (\pi^k_{\,k})^2 \right) \Bigg]
\nonumber\\[1ex]
& - \left[ \pi^{kj} D_k N^i + \pi^{ki} D_k N^j - D_k (\pi^{ij} N^k) \right]
+ \frac{8\pi G}{c^4} N \sum_{A = 1}^\mathcal{N}  \frac{p_{Ak}p_{Al}}{m_A}
\gamma^{ik}\gamma^{jl}
\left(1+\frac{p_{Am}p_{An}}{m_A^2 c^2}\gamma^{mn}\right)^{-\frac{1}{2}}
\delta_A \,.
\end{align}
In the ADM gauge, the trace (with respect to flat metric $\delta_{ij}$)
of \eqref{eq:momentum_evolution} reduces to
\begin{multline}
\label{eq:momentum_evolution_trace}
\Psi^2 \bigg(
\frac{d-2}{2} N \mathrm{R} - (d-1) \Psi^{-\frac{4}{d-2}} D_i D_i N \bigg)
- \frac{4-d}{2} \Psi^{-\frac{2(d+2)}{d-2}} N \pi^i_{\,j}\pi^j_{\,i}
- 2 \pi^{kl} D_k N^l
\\[1ex]
+ \frac{8\pi G}{c^4} N \Psi^{-\frac{8}{d-2}} \sum_{A = 1}^\mathcal{N}
       \frac{p_{A}^2}{m_A} 
\left(1+\frac{p_A^2}{m_A^2 c^2}\Psi^{-\frac{4}{d-2}}\right)^{-\frac{1}{2}}
       \delta_A = 0 \,. 
\end{multline}
Making now use of Eqs.\ \eqref{eq:curvature_density},
\eqref{eq:constraint_energy}, \eqref{eq:equation_Ni}, and of the two useful
relations $D_{(i} D_{j)} N = \gamma_{k(i} \partial_{j)} (D^k N) + \frac{1}{2}
\partial_k \gamma_{ij} D^k N$ and $D_i D_i N = 2 \partial_i N \partial_i \ln
\Psi + \Delta N$, Eq.\ \eqref{eq:momentum_evolution_trace} can be rewritten as
\begin{multline}
\label{eq:equation_N}
-(d-1) \Delta(N\Psi) + \frac{3d-2}{4} \Psi^{-\frac{3d-2}{d-2}} N
\pi^i_{\,j}\pi^j_{\,i} + \frac{4\pi G}{c^2} (d-2) \Psi^{-1} N
\sum_{A=1}^\mathcal{N} m_A
\left(1+\frac{p_A^2}{m_A^2 c^2}\Psi^{-\frac{4}{d-2}}\right)^{\frac{1}{2}}
\delta_A
\\[1ex]
+ \frac{8\pi G}{c^4} \Psi^{-\frac{d+2}{d-2}} N \sum_{A = 1}^\mathcal{N}
\frac{p_A^2}{m_A} \left(1+\frac{p_A^2}{m_A^2 
c^2}\Psi^{-\frac{4}{d-2}}\right)^{-\frac{1}{2}} \delta_A = 0 \,.
\end{multline}
The form of the latter equation suggests to solve it for the auxiliary function
$\chi\equiv{N\Psi}$ rather than for the lapse $N$ \cite{JS02}. At this
stage, we apply to the right-hand side of Eq.\ \eqref{eq:equation_N} the
replacement \eqref{eq:skeleton_approximation} which amounts to deleting a
flesh term in the product $\Psi^{-\frac{3d-2}{d-2}} \pi^i_{\,j}\pi^j_{\,i}$.
Since the Newtonian potential comes from the $1/c^2$ part of the lapse, the
term $\Psi^{-1} \Psi^{-\frac{3d-2}{d-2}} \pi^i_{\,j}\pi^j_{\,i} =
\mathcal{O}(1/c^6)$ is indeed a 2PN quantity and may be neglected consistently
with our approximation. Finally, we obtain a Poisson equation for
$\chi$ with the source term being some linear combination of the Dirac deltas:
\begin{multline}
\label{eq:equation_chi_skeleton}
\Delta\chi = \frac{4\pi G}{c^2} \frac{d-2}{d-1} \sum_{A=1}^\mathcal{N} \chi_A 
\Bigg( \frac{3d-2}{d-2} \frac{p_{Ai}V_{Ai}}{c} \Psi_A^{-\frac{4(d-1)}{d-2}}
\\[1ex]
+ m_A \Psi_A^{-2}
\bigg(1+\frac{p_A^2}{m_A^2c^2}\Psi_A^{-\frac{4}{d-2}}\bigg)^{-\frac{1}{2}} 
\left( 1 + \frac{d}{d-2}\frac{p_A^2}{m_A^2c^2}\Psi_A^{-\frac{4}{d-2}} \right)
\Bigg) \delta_A \,. 
\end{multline}
\end{widetext}
Following the same argument as in the derivation of the conformal factor in
Sec. \ref{sec:skeleton}, we may write 
\begin{equation}
\label{eq:chi}
\chi = 1 - A_d \sum_{A = 1}^\mathcal{N} \frac{\beta_A}{r_A^{d - 2}} \,,
\end{equation}
so that
$\Delta \chi = \frac{4 \pi G}{c^2} \frac{(d - 2)}{(d - 1)} \sum_{A =
  1}^\mathcal{N} \beta_A \delta_A$. We next replace the left-hand side of Eq.
\eqref{eq:equation_chi_skeleton} by virtue of the latter relation, and equate
the coefficients of $\delta_A$. We are led to
\begin{align}
  \beta_A & = \bigg( 1 - A_d \sum_{B \neq A} \frac{\beta_B}{r_{AB}^{d - 2}}
  \bigg) \Bigg( \frac{3 d - 2}{d - 2} \frac{p_{Ai} V_{Ai}}{c} \Psi_A^{-\frac{4
      (d - 1)}{d - 2}}
  \nonumber\\[1ex]
  &\quad + m_A \Psi_A^{-2} \left(1+\frac{p_A^2}{m_A^2c^2} \Psi_A^{-\frac{4}{d
        - 2}}\right)^{-\frac{1}{2}}
  \nonumber\\[1ex]
  & \qquad \times \left(1+ \frac{d}{d - 2} \frac{p_A^2}{m_A^2c^2}
    \Psi_A^{-\frac{4}{d - 2}} \right) \Bigg), \quad
  A=1,\,\ldots,\,\mathcal{N}.
\end{align}
This system of $\mathcal{N}$ equations for $\mathcal{N}$ unknowns provides in
principle the values of the monopoles $\beta_A$ entering the expression of
$\chi$, Eq.\ \eqref{eq:chi}. The lapse function itself is then given by
\[
N = \frac{\chi}{\Psi} \,.
\]
For $\mathcal{N}=2$ and $d=3$, we recover the result of Ref.\ \cite{JS02} in
the time-symmetric limit $p_{Ai}=0$ (with $A=1,2$).

In order to solve Eq.\ \eqref{eq:equation_Ni} for the shift function $N^i$, we
take the first and second order space derivatives of both sides. We obtain:
\begin{subequations}
\begin{align}
\label{eq:equation_Ni_explicit}
\Delta N^i + \frac{d-2}{d} \partial_{ij}N^j & = -2 \partial_{j}
\left(\Psi^{-\frac{2d}{d-2}} N \pi^i_{\,j} \right) \,,
\\[1ex]
\label{eq:equation_divergence_Ni_Wilson}
\frac{d-1}{d} \Delta(\partial_i N^i) & = - \partial_{ij}
\left(\Psi^{-\frac{2d}{d-2}} N \pi^i_{\, j} \right) \,. 
\end{align}
\end{subequations}
The source term of the second equation is the sum of a rational function of 
$r_A$, $A=1,\,\ldots,\,\mathcal{N}$, namely
$-\partial_i\big(\partial_j(\Psi^{-\frac{2d}{d-2}}N)\pi^i_{\,j}\big)$, plus a
distribution,
$-\partial_i\big(\Psi^{-\frac{2d}{d-2}}N\partial_j\pi^i_{\,j}\big)$.  The
Poisson integral of the term with non-compact support is not explicitly known,
but that of the distributional term is easy to compute. Now, the new ``flesh''
term does not contribute at the Newtonian level. It is therefore at least
post-Newtonian and will eventually play a role in the dynamics only
at the 2PN order; it will be consistently
neglected in the present model. Thus, the substitution
\eqref{eq:skeleton_approximation} will be supplemented by the replacement:
\begin{multline}
\label{eq:skeleton_approximation_Ni}
\partial_j \left( \Psi^{-\frac{2d}{d-2}} N  \pi^j_{\, i} \right)
\to \Psi^{-\frac{2d}{d-2}} N  \partial_j \pi^j_{\,i}
\\[1ex]
= -\frac{8\pi G}{c^3} \Psi^{-\frac{2d}{d-2}} N \sum_{A=1}^\mathcal{N} p_{Ai}
\delta_A \,.
\end{multline}

Taking Eq.\ \eqref{eq:skeleton_approximation_Ni} into account, Eq.\ 
\eqref{eq:equation_divergence_Ni_Wilson} becomes
\begin{equation}
\label{eq:equation_divergence_Ni}
\Delta(\partial_i N^i) = \frac{8\pi G}{c^3} \frac{d}{d-1} \partial_i
\bigg( \sum_{A=1}^\mathcal{N} \chi_A \Psi_A^{-\frac{3d-2}{d-2}}
p_{Ai} \delta_A \bigg) \,.
\end{equation}
The solution of \eqref{eq:equation_divergence_Ni} can be obtained
with the help of formula \eqref{eq:Laplacien_rd_2}. It reads
\begin{align}
\label{eq:divergence_Ni}
\partial_i N^i = & -\frac{2G}{c^3}
\frac{d}{d-1} \frac{\Gamma(\frac{d-2}{2})}{\pi^{\frac{d-2}{2}}}
\nonumber\\[1ex]
& \times \sum_{A=1}^\mathcal{N} \chi_A \Psi_A^{-\frac{3d-2}{d-2}}
p_{Ai} \partial_i \left( \frac{1}{r_A^{d-2}} \right) \,.
\end{align}
The Poisson
equation \eqref{eq:equation_Ni_explicit} after inserting the above expression
for $\partial_j N^j$ and performing the substitution 
\eqref{eq:skeleton_approximation_Ni} takes the form
\begin{align}
\label{eq:equation_Ni_skeleton}
\Delta N^i = & \frac{2G}{c^3}
\frac{d-2}{d-1} \frac{\Gamma(\frac{d-2}{2})}{\pi^{\frac{d-2}{2}}}
\nonumber\\[1ex]
& \quad \times
\sum_{A=1}^\mathcal{N}
\chi_A \Psi_A^{-\frac{3d-2}{d-2}}
p_{Aj}\partial_{ij}\left(\frac{1}{r_A^{d-2}}\right) 
\nonumber\\[1ex]
& + \frac{16\pi G}{c^3} \sum_{A=1}^\mathcal{N} \chi_A
\Psi_A^{-\frac{3d-2}{d-2}} p_{Ai} \delta_A \,.
\end{align}
By virtue of the formulas \eqref{eq:Laplacien_rm} and
\eqref{eq:Laplacien_rd_2}, the solution of Eq.\ 
\eqref{eq:equation_Ni_skeleton} is simply
\begin{multline}
\label{eq:Ni}
N^i = \frac{G}{c^3} \frac{\Gamma(\frac{d-2}{2})}{\pi^{\frac{d-2}{2}}}
\sum_{A=1}^\mathcal{N} \chi_A \Psi_A^{-\frac{3d-2}{d-2}}
\\[1ex]
\times \bigg( \frac{d-2}{(4-d)(d-1)} p_{Aj} \partial_{ij} r_A^{4-d}
- 4 \frac{p_{Ai}}{r_A^{d-2}} \bigg) \,.
\end{multline}

\section{Black-hole binary equations of motion}

For a black-hole binary, the number of coupled equations composing the system
\eqref{eq:alpha_A} that gives the effective masses reduces to two, and the sum
$\sum_{C \neq A}\alpha_C/r^{d-2}_{AC}$ contains only one term. It is now
convenient to regard the two quantities $\Psi_A$ as the unknowns:
\begin{multline}
\Psi_A = 1 + A_d \sum_{C \neq A} \frac{\alpha_C}{r^{d-2}_{AC}}
= 1 + A_d \frac{\alpha_B}{r^{d-2}_{12}},
\\[1ex]
A = 1,2, \quad B \neq A.
\end{multline}
Let us restrict ourselves to the space dimension $d= 3$. The constant
$A_d$ reads then $A_3=G/(2c^2)$, and the system \eqref{eq:alpha_A} consists of
two equations for the two unknowns $\Psi_B$: 
\begin{multline}
\label{eq:psi_A}
\Psi_B = 1 +  \frac{G m_A}{2 r_{12} c^2 \Psi_A}
\left( 1 + \frac{p_A^2}{m_A^2 c^2 \Psi_A^4} \right)^\frac{1}{2}
+ \frac{G p_{Ai} V_{Ai}}{2 r_{12} c^3 \Psi_A^7},
\\[2ex]
\quad B=1,2, \quad A \neq B \, .
\end{multline}
It is clearly possible to decouple the two equations \eqref{eq:psi_A}. In
Appendix \ref{app:polynomial} we show that each $\Psi_B$ is a root of a
polynomial of order 200.

The equations of motion of a black-hole binary read
\begin{subequations}
\label{eq:generic_equations_of_motion}
\begin{align}
\dot{x}^i_{A} & = \frac{\partial H}{\partial p_{Ai}},
\\[1ex]
\dot{p}_{Ai} & = -\frac{\partial H}{\partial x_{A}^i},
\end{align}
\end{subequations}
where the skeleton Hamiltonian $H$ expressed in terms of the quantities
$\Psi_A$ equals
\begin{equation}
\label{eq:skeleton_Hamiltonian_psi}
H = \frac{2 c^4 r_{12}}{G} (\Psi_1 + \Psi_2 - 2).
\end{equation}

Using the theorem of implicit differentiation, we put Eqs.\ 
\eqref{eq:generic_equations_of_motion} in a form apparently more convenient for
numerical integration. We start by differentiating Eq.\ \eqref{eq:psi_A} with
respect to the positions. We find (here $B=1,2$ and $A \neq B$):
\begin{subequations}
\begin{align}
\label{eq:psi_derivative_1_equation}
\partial_{1i}\Psi_B & \equiv \frac{\partial\Psi_B}{\partial x_1^i}
= \tilde{\eta}_{A1}^i - \zeta_A \partial_{1i} \Psi_A,
\\[1ex]
\label{eq:psi_derivative_2_equation}
\partial_{2i}\Psi_B & \equiv \frac{\partial\Psi_B}{\partial x_2^i}
= \tilde{\eta}_{A2}^i - \zeta_A \partial_{2i} \Psi_A,
\end{align}
\end{subequations}
where
\begin{subequations}
\label{eq:tilde_variables}
\begin{align}
\label{eq:eta_A1}
\tilde{\eta}_{A1}^i \equiv & -\frac{G m_A n_{12}^i}{2 c^2 r_{12}^2 \Psi_A}
\bigg( 1 + \frac{p_A^2}{m_A^2 c^2 \Psi_A^4} \bigg)^{\frac{1}{2}}
\nonumber\\[1ex]
& + \frac{G \partial_{1i} (p_{Aj} V_{Aj}/r_{12})}{2 c^3 \Psi_A^7},
\\[1ex]
\label{eq:eta_A2}
\tilde{\eta}_{A2}^i \equiv & -\tilde{\eta}_{A1}^i,
\\[1ex]
\label{eq:zeta_A}
\zeta_A \equiv & -\frac{\partial \Psi_B(x^i_C,p_{Ci},\Psi_C)}{\partial \Psi_A}
\nonumber\\[1ex]
= & \frac{G m_A}{2 c^2 r_{12} \Psi_A^2}
\frac{1 + 3 p_A^2/(m_A^2 c^2 \Psi_A^4)}
     {[1 + p_A^2/(m_A^2 c^2 \Psi_A^4)]^{\frac{1}{2}}}
\nonumber\\[1ex]
& + \frac{7 G \, p_{Aj} V_{Aj}}{2 c^3 r_{12} \Psi_A^8} \nonumber \\[1ex]
= & \frac{\chi_B - 1}{\chi_A}.
\end{align}
\end{subequations}
Equality \eqref{eq:eta_A2} follows from the fact that the right-hand side of
relation \eqref{eq:psi_A} depends on $x^i_1$ and $x^i_2$ only through the
black-hole separation $r_{12}\equiv[(x^i_1-x^i_2)(x^i_1-x^i_2)]^\frac{1}{2}$,
and that we have $\partial_{1i}r_{12}=-\partial_{2i}r_{12}$. Solving Eqs.\ 
\eqref{eq:psi_derivative_1_equation} with respect to $\partial_{1i}\Psi_B$ and
Eqs.\ \eqref{eq:psi_derivative_2_equation} with respect to
$\partial_{2i}\Psi_B$, we find
\begin{subequations}
\label{eq:psi_derivatives}
\begin{align}
\partial_{1i}\Psi_B &
= \frac{\tilde{\eta}^i_{A1}- \zeta_A \tilde{\eta}^i_{B1}}{1 - \zeta_1 \zeta_2},
\\[1ex]
\partial_{2i}\Psi_B & = -\partial_{1i}\Psi_B \, .
\end{align}
\end{subequations}

It remains to evaluate the momentum derivatives
$\partial_{p_{1i}}\Psi_B\equiv\partial\Psi_B/\partial p_{1i}$ and
$\partial_{p_{2i}}\Psi_B\equiv\partial\Psi_B/\partial p_{2i}$. The system of 
equations for $\partial_{p_{1i}}\Psi_B$ is given by
\begin{equation}
\label{eq:psi_derivative_p1_equation}
\partial_{p_{1i}}\Psi_B
= \tilde{\theta}_{A1}^i - \zeta_A \partial_{p_{1i}}\Psi_A,
\quad B=1,2, \quad A \neq B,
\end{equation}
where we have posed:
\begin{align}
\label{eq:theta_A1}
\tilde{\theta}_{A1}^i & \equiv \frac{G}{2 m_A c^4 r_{12} \Psi_A^5}
\frac{p_{1i} \delta_{A1}} {\displaystyle \left(1 + \frac{p_A^2}{m_A^2 c^2
      \Psi_A^4}\right)^{\frac{1}{2}}}
\nonumber\\[1ex]
& \quad + \frac{G\,\partial_{p_{1i}}(p_{Aj}V_{Aj})}{2 c^3 r_{12} \Psi_A^7}.
\end{align}
The coefficients $\zeta_A$ are identical to those appearing in
$\partial_{1i}\Psi_B$ because they represent the result of the
differentiation of the right-hand side of relation \eqref{eq:psi_A} with
respect to $\Psi_A$ at $x^i_A$ and $p_{Ai}$ constant.
The solution of Eqs.\ \eqref{eq:psi_derivative_p1_equation} with respect to
$\partial_{p_{1i}}\Psi_B$ reads
\begin{equation}
\label{eq:psi_derivative_p1}
\partial_{p_{1i}}\Psi_B = \frac{\tilde{\theta}^i_{A1}
- \zeta_A \tilde{\theta}^i_{B1}}{1 - \zeta_1 \zeta_2},
\end{equation}
from which we deduce $\partial_{p_{2i}}\Psi_B$ by exchanging the particle 
labels.

It is convenient to introduce the rescaled variables
\begin{align}
\label{eq:rescaled_variables}
\eta^i_A \equiv \frac{2 c^2 r_{12}}{G} \, \tilde{\eta}^i_{AA}, \quad
\theta^i_{AB} \equiv \frac{2 c^4 r_{12}}{G} \, \tilde{\theta}^i_{AB},
\quad A=1,2.
\end{align}
Making use of Eqs.\ \eqref{eq:skeleton_Hamiltonian_psi},
\eqref{eq:psi_derivatives}, \eqref{eq:psi_derivative_p1}, and
\eqref{eq:rescaled_variables}, the equations of motion
\eqref{eq:generic_equations_of_motion} can be written in the form
\begin{subequations}
\label{eq:equations_of_motion}
\begin{align}
  \dot{x}_1^i & = \frac{1}{1 - \zeta_1\zeta_2} \Big( (1 - \zeta_2)
  \theta_{11}^i + (1 - \zeta_1) \theta_{21}^i \Big) \,,
  \\[1ex]
  \dot{p}_{1i} & = -\frac{c^2}{1 - \zeta_1\zeta_2} \Big( (1 - \zeta_2)
  \eta_1^i - (1 - \zeta_1) \eta_2^i \Big)
  \nonumber\\[1ex]
  & \quad - \frac{2 c^4}{G} (\Psi_1 + \Psi_2 - 2) n_{12}^i \nonumber \\[1ex]
  & = \frac{c^2}{1 - \zeta_1 \zeta_2} \Big((1 - \zeta_1) \zeta_2 \eta_1^i - (1
  - \zeta_2) \zeta_1 \eta_2^i \Big) \nonumber \\[1ex] & \quad - c
  \left(\frac{1}{\Psi_1^7} + \frac{1}{\Psi_2^7}\right) \partial_{1i} (p_{1j}
  V_{1j}) \,,
\end{align}
\end{subequations}
where the last equality follows from relations \eqref{eq:psi_A} and definitions
\eqref{eq:eta_A1}--\eqref{eq:eta_A2}. Similar equations for $\dot{x}^i_2$ and
$\dot{p}_{2i}$ hold with the role of labels 1 and 2 exchanged. Knowing the
positions and momenta at a given time $t$, we can solve the algebraic
equations \eqref{eq:psi_A} numerically, calculate $\eta^i_A$, $\theta_{AB}^i$
and $\zeta_A$ by means of Eqs.\ \eqref{eq:tilde_variables},
\eqref{eq:theta_A1}, \eqref{eq:rescaled_variables}, and determine
$x_{A}^i(t+dt)$ as well as $p_{Ai}(t+dt)$ from the evolution equations
\eqref{eq:equations_of_motion}.

\section{Last stable circular orbit}

As an example of application of our skeleton Hamiltonian, we shall discuss in
this section the parameters of the last stable (circular) orbit (LSO) in the
relative motion of two inspiraling compact objects.  This class of systems is
of primary importance, being one of the most promising sources of detectable
gravitational waves for ground-interferometry experiments as LIGO (Laser
Interferometer Gravitational Wave Observatory), VIRGO, GEO600, and TAMA300.
Because of gravitational-wave emission, the separation between the bodies
decreases adiabatically in time while the frequency increases according to the
relativistic version of the Kepler law.  The process ends as soon as the first
instable orbit is reached before the coalescence phase.  The point where this
occurs changes according to the model (see Refs.\ \cite{DJS00c,B02} and
references therein for some comparisons).

For binary black holes the skeleton Hamiltonian reads
$H=(\alpha_1+\alpha_2)c^2$, where $\alpha_1$ and $\alpha_2$ are solutions of
the coupled system of equations \eqref{eq:alpha_A}.  It is impossible to solve
this system analytically, but it is not difficult to solve it perturbatively
within the post-Newtonian setting, treating $\varepsilon\equiv{1/c^2}$ as a
small parameter.  At the $n$PN approximation, we search the post-Newtonian
expansion of the coefficients $\alpha_A$ in the form
\begin{equation}
\label{eq:alpha_A_PN}
\alpha_A = m_A + \sum_{i=1}^{n+1} \alpha_A^{(i-1)} \varepsilon^{i}
+ \mathcal{O}(\varepsilon^{n+2}) \,, \quad A=1,2 \,.
\end{equation}
Note that we need to carry the series up to the order $n+1$ in $\varepsilon$
to get an $n$PN-accurate dynamics with respect to the exact skeleton evolution
(which first differs from the exact Einstein's evolution at the 2PN order). We
now substitute \eqref{eq:alpha_A_PN} into Eqs.\ \eqref{eq:alpha_A} and expand
their right-hand sides up to the order $n+1$. By identifying the coefficients
of different powers $\varepsilon^i$, $i=1,\,\ldots,\,n+1$, we obtain a system
of algebraic equations for $\alpha_A^{(i)}$. We display here the first three
equations:
\begin{align*}
\frac{\alpha_1^{(0)}}{m_1} &=
\frac{p_{1i} p_{1i}}{2 m_1^2} - \frac{G m_2}{2r_{12}},
\\[2ex]
\frac{\alpha_1^{(1)}}{m_1} &=
- \frac{(p_{1i}p_{1i})^2}{8m_1^4} + \frac{G m_2}{2r_{12}} \left(
- \frac{5 p_{1i} p_{1i}}{2 m_1^2}
+ \frac{7 p_{1i} p_{2i}}{2 m_1 m_2} \right.
\\[2ex]
& \quad \left. + \frac{n_{12}^i n_{12}^j p_{1i} p_{2j}}{2 m_1 m_2}
- \frac{\alpha_2^{(0)}}{m_2} \right) + \frac{G^2 m_2^2}{4 r_{12}^2},
\\[2ex]
\frac{\alpha_1^{(2)}}{m_1} &=
\frac{(p_{1i} p_{1i})^3}{16 m_1^6}
+ \frac{G m_2}{2 r_{12}} \left(\frac{9 (p_{1i}p_{1i})^2}{8 m_1^4}
- \frac{5 p_{1i}p_{1i} \alpha_2^{(0)}}{2 m_1^2 m_2} \right.
\\[2ex]
& \quad \left. - \frac{\alpha_2^{(1)}}{m_2} \right)
+ \frac{G^2 m_2^2}{2 r_{12}^2} \left(
\frac{15 p_{1i} p_{1i}}{4 m_1^2}
- \frac{49 p_{1i} p_{2i}}{4 m_1 m_2} \right.
\\[2ex]
& \quad \left. - \frac{7 n_{12}^i n_{12}^j p_{1i} p_{2j}}{4 m_1 m_2}
+ \frac{\alpha_2^{(0)}}{m_2} \right)
- \frac{G^3 m_2^3}{8 r_{12}^3} \,;
\end{align*}
similar equalities hold with labels 1 and 2 exchanged.
It is straightforward to solve up this system to the required order. 

For convenience, we shall consider instead of $H$ the dimensionless
Hamiltonian (after ruling out the constant rest-mass contribution)
$$
\hat{H} \equiv \frac{H - (m_1+m_2) c^2}{\mu c^2},
$$
$\mu\equiv m_1 m_2/(m_1 + m_2)$ denoting the reduced mass of the
system. At the $n$PN level, we have
\begin{equation}
\label{eq:H_PN}
\hat{H} = \sum_{A=1}^2 \sum_{i=1}^{n+1} \frac{\alpha_A^{(i-1)}}{\mu} 
\varepsilon^{i} + \mathcal{O}\left(\varepsilon^{n+2}\right).
\end{equation}

We restrict our study to circular orbits in the center-of-mass reference
frame. Within the ADM gauge, the frame shift is achieved by imposing the
condition
\begin{equation}
\label{eq:center_of_mass}
p_{1i} + p_{2i} = 0 \, ,
\end{equation}
whereas circularity requires that
\begin{equation}
\label{eq:circular_orbit}
n^i_{12} p_{1i} = n^i_{12} p_{2i} = 0 \, .
\end{equation}
The form of $\hat{H}$ becomes particularly simple when use is made
of the reduced relative position and momentum:
$$
r^i \equiv \frac{x^i_1-x^i_2}{G(m_1+m_2)} \,, \quad
p_i \equiv \frac{p_{1i}}{\mu} = -\frac{p_{2i}}{\mu} \,.
$$
When both relations \eqref{eq:center_of_mass} and \eqref{eq:circular_orbit}
hold, the Hamiltonian \eqref{eq:H_PN} can be expressed with the help of
$r^i$ and $p_i$ only. Notice also that it depends on the masses
$m_1$ and $m_2$ exclusively through the symmetric mass ratio $\nu \equiv m_1
m_2/(m_1+m_2)^2$.

In the next step, we introduce the conserved (reduced) angular momentum $j$
along the cirular orbit of radius $r$:
\begin{equation}
\label{eq:pp}
p^2 = \frac{j^2}{r^2}.
\end{equation}
After eliminating the linear momentum $p_i$ by means of Eq.\ \eqref{eq:pp},
the dimensionless energy $E\equiv\hat{H}(r,j)$ of the binary system becomes a
function of the variables $r$ and $j$:
\begin{align}
\label{eq:E_rj}
\frac{E}{\varepsilon}
& = \frac{j^2}{2r^2} - \frac{1}{r}
+ \bigg( \frac{1}{r^2} - \frac{j^2}{r^3}(3+\nu) + \frac{j^4}{4 r^4}(-1+3\nu)
\bigg) \frac{\varepsilon}{2}
\nonumber \\
& \quad + \bigg( -\frac{1}{2r^3} (1+\nu) + \frac{j^2}{r^4} (5+\nu)
+ \frac{5j^4}{4r^5} (1-4\nu)
\nonumber \\
& \quad\quad\quad + \frac{j^6}{8 r^6} (1 - 5\nu + 5\nu^2) \bigg)
\frac{\varepsilon^2}{2}
+ {\cal O}(\varepsilon^3).
\end{align}

The coordinate radius $r$ of the circular orbit and the angular momentum $j$
are related through one of the canonical equations of motion, namely
\begin{equation}
\label{eq:circularity_condition}
\frac{\partial\hat{H}(r,j)}{\partial r} = 0 \,,
\end{equation}
a relation that may be regarded as expressing the quasi-stationarity of the
configuration.  Remarkably, it can be shown in this case, i.e., for circular
motion, that the ADM mass of the system equates the so-called Komar mass
$M_\text{K}$, i.e., the mass-like quantity proportional to the
monopole part of $g_{00}$ at spatial infinity satisfying the relation
$g_{00}=-1+4A_3M_\text{K}/r+\mathcal{O}(1/r^2)$ \cite{K59}. The proof of the
equality $M_\text{ADM} = M_\text{K}$ is given in Appendix \ref{app:komar} for
an arbitrary number of space dimensions.

Equation \eqref{eq:circularity_condition} is solved perturbatively for
$r=r(j)$.  By plugging the result into Eq.\ \eqref{eq:E_rj}, we find the gauge
invariant link between the energy $E$ and the angular momentum $j$ in the case
of circular motion:
\begin{align}
\label{eq:E_j}
\frac{E}{\varepsilon}
= & \frac{\hat{H}(r(j),j)}{\varepsilon}
\nonumber\\[1ex]
= & -\frac{1}{2j^2} - \frac{1}{8j^4} (9+\nu) \varepsilon
\nonumber\\[1ex]
& - \frac{1}{16j^6} \big(81+41\nu-5\nu^2\big) \varepsilon^2 + {\cal
O}(\varepsilon^3).
\end{align}

It is convenient to rewrite formula \eqref{eq:E_j} in terms of the
dimensionless variable
\begin{equation}
\label{eq:x}
x \equiv \left(\sqrt{\varepsilon}\,\frac{dE}{dj}\right)^{\frac{2}{3}}.
\end{equation}
The parameter $x$ is now treated as a small quantity of the order $\varepsilon$
[cf.\ Eq.\ \eqref{eq:E_j}]. The definition \eqref{eq:x} establishes the link
between $x$ and $j$. We invert it order by order to obtain $j$ as a
function of $x$. Inserting $j=j(x)$ into Eq.\ \eqref{eq:E_j} leads to a gauge
invariant expression of the energy $E$ depending on the parameter $x$. The
calculation is performed at the 10PN approximation in order to come
reasonably close to the exact result. We display here the explicit formula
with the 3PN-accuracy. It reads
\begin{align}
\label{eq:E_x}
E = & -\frac{x}{2} + \bigg(\frac{3}{8}+\frac{\nu}{24}\bigg) x^2
+ \bigg(\frac{27}{16}+\frac{29}{16}\nu-\frac{17}{48}\nu^2\bigg) x^3
\nonumber\\
& + \bigg(\frac{675}{128}+\frac{8585}{384}\nu-\frac{7985}{192}\nu^2
+\frac{1115}{10368}\nu^3\bigg) x^4
\nonumber\\
& + \sum_{i=5}^{11} e_i x^i + \mathcal{O}(x^{12}).
\end{align}
The higher order post-Newtonian coefficients $e_i$ (for $i=5,\,\ldots,\,11$)
can be found in Appendix \ref{app:10PN}.

We want to compare the result of Eq.\ \eqref{eq:E_x} with the relation
$E_\text{WM}=E_\text{WM}(x)$ of the Wilson-Mathews model (remember that
in this model the post-Newtonian expansion terminates at the 2PN
order)\footnote{This relation is obtained by taking $n=j$ in Eq.\ (C4) of
  Ref.\ \cite{DJS00a} and replacing there the angular momentum $j$ by
  $j_\text{WM}(x)$ according to Eq.\ (B1) of Ref.\ \cite{DJS00c}.}:
\begin{align}
E_\text{WM}
= & -\frac{x}{2} + \bigg( \frac{3}{8} + \frac{\nu}{24} \bigg) x^2
\nonumber\\
& + \bigg( \frac{27}{16} - \frac{39}{16}\nu - \frac{17}{48} \nu^2 \bigg) x^3
+ \mathcal{O}(x^4).
\end{align}
The 2PN coefficient proportional to $\nu^2$ differs from that of the
post-Newtonian computations carried out in general relativity. Indeed, the
energy for circular orbits in Einsteinian theory, up to the 3PN order, is
\cite{DJS00c}
\begin{align}
\label{eq:EGR}
E_\text{GR}
= & -\frac{x}{2} + \bigg( \frac{3}{8} + \frac{\nu}{24} \bigg) x^2
+ \bigg( \frac{27}{16} - \frac{19}{16}\nu + \frac{1}{48}\nu^2 \bigg) x^3
\nonumber\\
& + \Bigg( \frac{675}{128} +
\bigg(\frac{205}{192}\pi^2-\frac{34445}{1152}\bigg)\nu
\nonumber\\
& \quad + \frac{155}{192} \nu^2 + \frac{35}{10368} \nu^3 \Bigg) x^4
+ \mathcal{O}(x^5),
\end{align}
where, according to Ref.\ \cite{DJS01b} we set the parameter
$\omega_\text{static}$ (defined in paper \cite{JS00}) equal to zero
($\omega_\text{static}=0$), getting the solution corresponding to the
Brill-Lindquist initial data.\footnote{If the space transformation which
  relates the Brill-Lindquist solution with the Misner-Lindquist one through
  3PN order \cite{JS99} is regarded as induced by canonical transformation,
  the Brill-Lindquist and Misner-Lindquist solutions are physically identical
  at this level.} By comparing Eqs.\ \eqref{eq:E_x} and \eqref{eq:EGR} it is
obvious that our energy $E$ starts to differ from the general-relativistic one,
$E_\text{GR}$, at the 2PN order.

\begin{figure}
\resizebox{8.5cm}{!}{\includegraphics{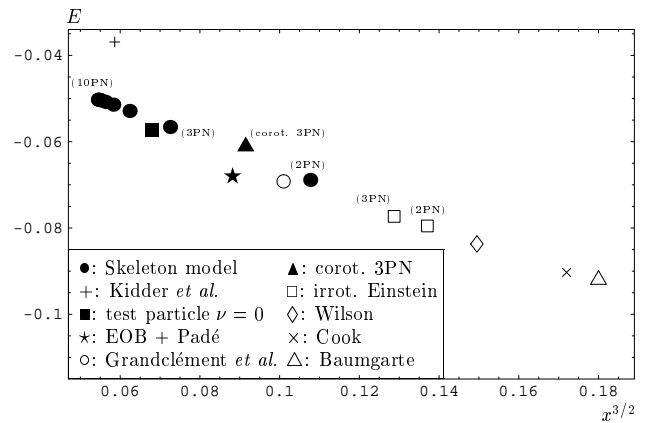}}
\caption{\label{fig:Ex_plot}
  Position of the last stable circular orbit in the frequency/energy plane for
  equal-mass binaries in different models.  The sequence of black circles
  corresponds to different post-Newtonian truncations of the skeleton energy
  (from the 2PN to the 10PN order when moving to the left).  The values of the
  3PN Einsteinian energy for non-spinning black holes are computed in Ref.\ 
  \cite{B02} based on the works of Damour, Jaranowski, Sch{\"a}fer
  \cite{DJS00b,DJS01e00,DJS01,DJS01b} and Blanchet, Faye
  \cite{BF00a,BF01b,ABF01} with $\omega_\text{static}=0$.  The modifications
  coming from the spins in the corotational case are given in Ref.\ 
  \cite{B02}.  The plus cross refers to the point obtained within the hybrid
  method of Kidder, Will, and Wiseman \cite{KWW92a}, in which terms are added
  to the 2PN equations of motion in order to get the exact test-particle
  limit.  The data corresponding to the white circle have been calculated from
  a sequence of quasi-equilibrium configurations for corotating Misner black
  holes by Grandcl{\'e}ment and collaborators \cite{GGB01a,GGB01b} using a
  Wilson-Mathews-type approximation (conformal thin-sandwich approach
  \cite{C03}).  The time cross indicates the values computed by Cook
  \cite{C94} for two Misner black holes assuming a Bowen-York-like extrinsic
  curvature (conformal transverse-traceless approach \cite{C03}).  The white
  triangle shows the position of the last stable circular orbit determined by
  Baumgarte \cite{B00} with the help of the puncture method, which means in
  particular that the black holes are of Brill-Lindquist type, and that the
  extrinsic curvature is the one of Bowen-York (conformal transverse-traceless
  approach).  The star refers to the effective-one-body approach with Pad{\'e}
  approximant \cite{DJS00c}.  Finally, we have indicated the frequency and the
  energy in the test-body limit ($\nu=0$).}
\end{figure}

\begin{table*}
\caption{\label{tab:Ex_table}
The last stable circular orbit parameters in the skeleton model
truncated at various PN orders for equal-mass binaries.}
\begin{ruledtabular}
\begin{tabular}{cdddddddddd}
& \text{1PN} & \text{2PN} & \text{3PN} & \text{4PN} & \text{5PN}
& \text{6PN} & \text{7PN} & \text{8PN} & \text{9PN} & \text{10PN}
\\ \hline $x^{3/2}$
& 0.5224
& 0.1077
& 0.0725
& 0.0624
& 0.0583
& 0.0563
& 0.0554
& 0.0548
& 0.0546
& 0.0544
\\ $E$
& -0.1622
& -0.0689
& -0.0566
& -0.0529
& -0.0514
& -0.0508
& -0.0505
& -0.0503
& -0.0503
& -0.0502
\end{tabular}
\end{ruledtabular}
\end{table*}

The position $x_\text{LSO}$ of the last stable orbit is determined from the
requirement that $(dE/dx)_{x=x_\text{LSO}}=0$.\footnote{This determination is
  not unique for approximate expressions.  The method presented in paper
  \cite{BI03} is also valuable.  It will not be employed here since we are
  interested in the comparison with other works.}  The corresponding energy
$E_\text{LSO}$ [computed by means of Eq.\ \eqref{eq:E_x}] is depicted in
Fig.~\ref{fig:Ex_plot} for equal-mass binaries (i.e., for $\nu=1/4$) as a
function of the dimensionless orbital frequency $x_\text{LSO}^{3/2}$.
Numerical values are given in Table~\ref{tab:Ex_table}. The case of a test
particle moving in a Schwarzschild background is also plotted for comparison.
Similarly to what happens in Einsteinian theory \cite{B02}, the instabilities
arise ``earlier'' (i.e., at lower frequency) for higher post-Newtonian orders
in the skeleton model. The 2PN last stable orbit of our skeleton dynamics is
quite far from that of general relativity in the sense that the absolute
difference $\delta (x^{3/2}_\text{LSO})_\text{2PN}$ ($\delta
(E_\text{LSO})_\text{2PN}$) between the skeleton frequency
$(x^{3/2}_\text{LSO})_\text{2PN} = 0.1077$ (the skeleton energy
$(E_\text{LSO})_\text{2PN} = -0.0689$) and the Einsteinian one
$(x^{3/2}_{\genfrac{}{}{0pt}{}{\text{GR}}{\text{LSO}}})_\text{2PN} = 0.1371$
($(E_{\genfrac{}{}{0pt}{}{\text{GR}}{\text{LSO}}})_\text{2PN} = -0.0795$) is
not negligible with respect to the general-relativistic value: $\delta
(x^{3/2}_\text{LSO})_\text{2PN}/
(x^{3/2}_{\genfrac{}{}{0pt}{}{\text{GR}}{\text{LSO}}})_\text{2PN} = 21$\%
($\delta (E_\text{LSO})_\text{2PN}/
(E_{\genfrac{}{}{0pt}{}{\text{GR}}{\text{LSO}}})_\text{2PN} = 13$\%). The
location of the Wilson-Mathews data suggests that modifications of the
extrinsic curvature affect the parameters of the last stable orbit more than
the choice of a conformally flat space metric \cite{PCT02}. The 3PN skeleton
values are even further away from the Einsteinian 3PN ones, but they are
closer to those of the effective one-body approach improved by Pad{\'e}
approximants than the straight 3PN Einsteinian values.  The 10PN data are
particularly interesting, considering that they may be regarded as good
approximants to the full skeleton model. Remarkably, their position in the
$(x^{3/2},E)$ plane relative to the 3PN data is nearly the same as that of the
point referring to numerical simulations relative to the point referring to
the 3PN data in general relativity.

In Fig.\ \ref{fig:Ex_plot}, the relative location of the last stable orbit
indicated by the white circle obtained from numerical computations with
respect to the 3PN result for corotating bodies (black triangle) is somehow
plausible if we believe that it is essentially due to the Wilson-Mathews
truncation.  It is indeed comparable to the relative location of irrotational
2PN Einsteinian and Wilson-Mathews data.  We also want to point out that a
realistic estimation of the last stable orbit does not necessarily follow from
the standard 3PN calculation.  It may well be close to the star-like point
derived from the effective one-body model with Pad\'e approximant.  There is
currently no definite argument to decide what is the most likely statement.
As a result, we regard the position of these points as defining the present
uncertainty in the location of the last stable orbit derived analytically.

\begin{acknowledgments}
  
  P.J.\ thanks the Theoretisch-Physikalisches Institut of the FSU Jena for
  hospitality during the realization of the present work. G.S.\ is thankful to
  the spring-2003 numerical relativity group at Caltech, particularly Greg
  Cook, Lee Lindblom, Niall {\'O} Murchadha, and Masaru Shibata, for
  stimulating discussions.  We also thank Christian Thierfelder for useful
  comments about the presentation of the evolution equations. This work is
  supported by the EU Programme ``Improving the Human Research Potential and
  the Socio-Economic Knowledge Base'' (Research Training Network Contract
  HPRN-CT-2000-00137), by the Caltech visitors program, and by the Deutsche
  Forschungsgemeinschaft (DFG) through SFB/TR7
  ``Gravitationswellenastronomie''. P.J.\ got an additional support from the
  Polish KBN Grant No.\ 5 P03B 034 20.

\end{acknowledgments}

\appendix

\section{Polynomial equations for the coefficients $\Psi_A$
in $d=3$ dimensions} \label{app:polynomial}

Equation \eqref{eq:psi_A} for $A=2$, after multiplication by $\Psi_1^7$, can
be written in the form
$\Psi_2 \Psi_1^7 = P_{1 \, (7)} + \Psi_1^4 P_{1 \,
(4)}^{1/2}$, where $P_{1\,(7)} \equiv \Psi_1^7 + G p_{1i} V_{1i}/(2 r_{12}
c^3)$ and $P_{1 \, (4)} \equiv [G m_1/(2 r_{12}c^2)]^2 [\Psi_1^4 + p_1^2/(m_1^2
c^2)]$ are polynomials in $\Psi_1$ with non-zero constant terms, of order 7
and 4, respectively. A similar relation holds for $A=1$. We now subtract
$P_{2 \, (7)}$ from the left and right-hand side, and square the new
identity. We obtain:
\begin{multline}
\label{eq:equation_gamma1}
(\Psi_1 - 1)^2 \Psi_2^{14}
- \frac{G p_{2i} V_{2i}}{r_{12} c^3} (\Psi_1 - 1) \Psi_2^7
+ \left(\frac{G p_{2i} V_{2i}}{2 r_{12} c^3}\right)^2
\\[1ex]
= \left(\frac{G m_2}{2 r_{12} c^2} \right)^2 \Psi_2^8
\left(\Psi_2^4 + \frac{p_2^2}{m_2^2 c^2}\right) \,.
\end{multline}

Next, we multiply the equality (\ref{eq:equation_gamma1}) by $(\Psi_1^7)^{14}=
\Psi_1^{98}$, make the substitution $\Psi_2 \to \Psi_1^{-7} (P_{1 \, (7)} +
\Psi_1^4 P_{1 \, (4)}^{1/2})$, and expand the terms $(P_{1 \, (7)} + \Psi_1^4
P_{1 \, (4)}^{1/2})^n$, $n\in \mathbb{N}^*$, in powers of $P_{1 \, (4)}^{1/2}$
with the help of the binomial formula. If $n$ is even, the contribution
$M_1^{(n)}\equiv\sum_{k=0}^{n/2}\genfrac{(}{)}{0pt}{}{n}{2k} P_{1\,(7)}^{2k}
\Psi_1^{8(n/2-k)}P_{1\,(4)}^{n/2-k}$, with
$\genfrac{(}{)}{0pt}{}{n}{2k}=n!/[(2k)!(n-2k)!]$, is of polynomial kind,
whereas that of $\sum_{k=0}^{n/2-1} \genfrac{(}{)}{0pt}{}{n}{2k+1}
P_{(7)}^{2k+1} \Psi_1^{8(n/2 -k - 1)+4} P_{1 \, (4)}^{n/2-k-1+1/2}$ has the
form of a polynomial $N_1^{(n)}$ multiplied by the algebraic function $P_{1 \,
  (4)}^{1/2}$; $M_1^{(n)}$ and $N_1^{(n)}$ have degrees $\max_{0\leq k \leq
  n/2} [7\times2k + 12\times(n/2-k)]= 7n$ and $\max_{0\leq k \leq n/2-1}
[7\times(2k+1) + 12\times(n/2-k-1) + 4] = 7n - 3$, respectively. The same
decomposition holds for odd $n$ with different expressions for $M_1$ and
$N_1$. At last, we gather all polynomial quantities into the first member of
the equality while the non polynomial contributions $P_{1\,(4)}^{1/2}$ are
recast to the second member. The final equation is
\begin{widetext}
\begin{align}
\label{eq:gamma1}
&(\Psi_1 - 1)^2 \sum_{k=0}^7 \genfrac{(}{)}{0pt}{}{14}{2k} P_{1 \, (7)}^{2k} 
\Psi_1^{8 (7 - k)} P_{1 \, (4)}^{7-k}  -
\frac{G p_{2i} V_{2i}}{r_{12} c^3} (\Psi_1 - 1) \Psi_1^{49}
\sum_{k=0}^3 \genfrac{(}{)}{0pt}{}{7}{2k+1} P_{1\, (7)}^{2k+1}
\Psi_1^{8(3-k)} P_{1\, (4)}^{3-k} 
+ \left(\frac{G p_{2i} V_{2i}}{2 r_{12} c^3} \right)^2 \Psi_1^{98}
\nonumber\\
& - \left(\frac{G m_2 \Psi_1^7}{2 r_{12} c^2} \right)^2
\left[ \sum_{k=0}^6 \genfrac{(}{)}{0pt}{}{12}{2k} P_{1\, (7)}^{2k}
  \Psi_1^{8(6-k)} P_{1\, 
(4)}^{6-k} + \frac{p_2^2 \Psi_1^{28}}{m_2^2 c^2}
\sum_{k=0}^{4} \genfrac{(}{)}{0pt}{}{8}{2k} P_{1\, (7)}^{2k} \Psi_1^{8(4-k)}
P_{1\, (4)}^{4-k} \right]
\nonumber\\
& = - P_{1 \, (4)}^{1/2} \Bigg[(\Psi_1 - 1)^2 \sum_{k=0}^6
\genfrac{(}{)}{0pt}{}{14}{2k +1} P_{1\, (7)}^{2k+1} \Psi_1^{4 (13 - 2k)}
P_{1\, (4)}^{6-k} - 
\frac{G p_{2i} V_{2i}}{r_{12} c^3} (\Psi_1 - 1) \Psi_1^{49}
\sum_{k=0}^3 \genfrac{(}{)}{0pt}{}{7}{2k} P_{1\, (7)}^{2k} \Psi_1^{4 (7 -
  2k)} P_{1\, (4)}^{3-k} 
\nonumber\\
& \qquad \qquad - \left(\frac{G m_2 \Psi_1^7}{2 r_{12} c^2} \right)^2
\Bigg(\sum_{k=0}^5 \genfrac{(}{)}{0pt}{}{12}{2k+1} P_{1 \, (7)}^{2k+1}
\Psi_1^{4 (11-2k)} P_{1 \, (4)}^{5-k} + \frac{p_2^2
\Psi_1^{28}}{m_2^2 c^2}  \sum_{k=0}^3 \genfrac{(}{)}{0pt}{}{8}{2k+1} P_{1\,
(7)}^{2k+1} 
\Psi_1^{4 (7-2k)} P_{1\, (4)}^{3-k} \Bigg) \Bigg] \,.
\end{align}
We see that the left-hand side of Eq.\ \eqref{eq:gamma1} is made of a sum of
polynomials of degrees 100, 99, 98, 98, 98, the first of them involving the
only non-zero constant term.  The right-hand side is the product of
$P_{1\,(4)}^{1/2}$ and polynomials of degrees 97, 96, 95, 95 respectively that
cancel when $\Psi_1=0$. As a consequence, $\Psi_1$ is the root of a polynomial
of degree 200 with a non-zero constant term.
\end{widetext}

\section{Proof of equality of ADM and Komar masses
in the case of two bodies along circular orbits} \label{app:komar}

In this appendix, we restrict ourselves to a binary system. The black
holes are thus labeled by $A,B,\ldots\in\{1,2\}$. Our starting point is the
system of two sets of equations satisfied by the monopoles $\alpha_A$ and
$\beta_A$ entering the conformal factor $\Psi$ and the auxiliary function
$\chi\equiv{N\Psi}$:
\begin{subequations}
\label{eq:alpha_beta_binary}
\begin{align}
\label{eq:alpha_A_binary}
\alpha_A & = \frac{p_{Ai}V_{Ai}}{c} \Psi_A^{-\frac{3d-2}{d-2}}
\nonumber\\[2ex]
& \quad + m_A \Psi_A^{-1}
\left(1+\frac{p_A^2}{m_A^2c^2}\Psi_A^{-\frac{4}{d-2}}\right)^{\frac{1}{2}} \,,
\\[2ex]
\label{eq:beta_A_binary}
\beta_A & = \chi_A \Bigg( \frac{3d-2}{d-2} \frac{p_{Ai}V_{Ai}}{c}
\Psi_A^{-\frac{4(d-1)}{d-2}}
\nonumber\\[2ex]
& \quad + m_A \Psi_A^{-2} 
\left(1+\frac{p_A^2}{m_A^2c^2}\Psi_A^{-\frac{4}{d-2}}\right)^{-\frac{1}{2}}
\nonumber\\[2ex]
& \qquad \times
\left(1 + \frac{d}{d-2} \frac{p_A^2}{m_A^2c^2} \Psi_A^{-\frac{4}{d-2}} \right) 
\Bigg) \,, 
\end{align}
\end{subequations}
with (here $B \neq A$)
\begin{subequations}
\label{eq:psi_chi_binary}
\begin{align}
\label{eq:psi_A_binary}
\Psi_A & = 1 + A_d \frac{\alpha_B}{r_{12}^{d - 2}} \,,
\\[2ex]
\label{eq:chi_A_binary}
\chi_A & = 1 - A_d \frac{\beta_B}{r_{12}^{d - 2}} \,.
\end{align}
\end{subequations}

It is straightforward to show that for a binary system described by the
functions $\Psi$ and $\chi$ of the purely monopolar form given by Eqs.\ 
\eqref{eq:psi} and \eqref{eq:chi}, the ADM mass coincides with
its Komar mass if and only if the quantities $\alpha_A$ and $\beta_A$ satisfy
the simple relation:
\begin{equation}
\label{eq:equivalence_ADM_Komar}
\sum_{A=1}^2 (\alpha_A - \beta_A) = 0 \,.
\end{equation}

We shall assume that the relative orbits are circular. After introducing the
reduced relative momentum $p_i$ and eliminating its square through the
relation $p^2=G^2 (m_1 + m_2)^2 j^2/r_{12}^2$, all quantities become functions
of $r_{12}$ and $j$ only. By virtue of Eqs.\ \eqref{eq:alpha_A_binary}, the
effective mass $\alpha_A$ can then be treated as a function of the {\em three}
variables $\Psi_A$, $r_{12}$, and $j$:
\begin{equation}
\label{eq:alpha_A_implicit}
\alpha_A = \alpha_A(\Psi_A(r_{12},j),r_{12},j)\, .
\end{equation}
With the help of Eqs.\ \eqref{eq:alpha_beta_binary}, it is not difficult to
check that the following relations, crucial for our proof, are fulfilled by
$\alpha_A$:
\begin{subequations}
\label{eq:alpha_A_derivative}
\begin{align}
\label{eq:alpha_A_derivative_psi_A}
\frac{\partial \alpha_A(\Psi_A,r_{12},j)}{\partial \Psi_A} &= -
\frac{\beta_A}{\chi_A}\, , 
\\[2ex]
\label{eq:alpha_A_derivative_r12}
\frac{\partial \alpha_A(\Psi_A, r_{12}, j)}{\partial r_{12}}
&= \frac{d - 2}{2r_{12}} \bigg(\alpha_A - \beta_A \frac{\Psi_A}{\chi_A}
\bigg)\, . 
\end{align}
\end{subequations}
Note that, writing down Eqs.\ \eqref{eq:alpha_A_derivative_r12}, we have
already employed the relations \eqref{eq:alpha_A_derivative_psi_A}.

The ADM Hamiltonian, restricted to circular orbits, reads
\begin{equation}
\label{eq:circular_orbit_Hamiltonian}
H(r_{12},j) = c^2 \sum_{A=1}^2 \alpha_A(r_{12},j).
\end{equation}
The circularity condition implies
$\partial H(r_{12},j)/\partial r_{12}=0$, which is equivalent to
\begin{equation}
\label{eq:circularity_condition_alpha}
\sum_{A=1}^2 
\frac{\partial \alpha_A(r_{12},j)}{\partial r_{12}} = 0.
\end{equation}
After substituting \eqref{eq:circular_orbit_Hamiltonian} into
\eqref{eq:circularity_condition_alpha} we obtain the
equality (valid along circular orbits): 
$$
\sum_{A=1}^2 \left(
\frac{\partial \alpha_A}{\partial \Psi_A}\frac{\partial \Psi_A}{\partial
  r_{12}} + \frac{\partial \alpha_A}{\partial r_{12}} \right) = 0. 
$$
According to Eq.\ \eqref{eq:alpha_A_derivative}, it can be rewritten as
\begin{equation}
\label{eq:improved_circularity_condition}
\sum_{A=1}^2 \Bigg( \frac{\beta_A}{\chi_A}
\frac{\partial\Psi_A}{\partial r_{12}} 
+ \frac{d - 2}{2r_{12}} \left(
\beta_A \frac{\Psi_A}{\chi_A} - \alpha_A \right) \Bigg) = 0 \,.
\end{equation}

On the other hand, the equations \eqref{eq:psi_A_binary} in the circular case
take the form
\[
\Psi_A(r_{12},j) = 1 + A_d 
\frac{\alpha_B(\Psi_B(r_{12},j),r_{12},j)}{r_{12}^{d - 2}}\,, \quad B \neq A
\,. 
\]
By differentiating both sides with respect to $r_{12}$ and resorting to
relations \eqref{eq:alpha_A_derivative}, we obtain the system of two equations
(where $B\neq A$)
\begin{equation*}
\frac{\partial \Psi_A}{\partial r_{12}} = -
\frac{d - 2}{2r_{12}^{d - 1}} A_d \left(\alpha_B +
 \beta_B \frac{\Psi_B}{\chi_B}   \right) 
- \frac{A_d}{r_{12}^{d - 2}} \frac{\beta_B}{\chi_B} \frac{\partial
  \Psi_B}{\partial   r_{12}} \, ,
\end{equation*}
which we solve with respect to the derivatives ${\partial\Psi_A}/{\partial
  r_{12}}$. After eliminating $\Psi_A$ and $\chi_A$ by means of relations
  \eqref{eq:psi_chi_binary}, the solution reduces to
\begin{equation}
\label{eq:equality_condition}
\frac{\partial \Psi_A}{\partial r_{12}}
= - \frac{d - 2}{2 r_{12}^{d - 1}} A_d (\alpha_B + \beta_B )\, ,
\quad B \neq A \, .
\end{equation}

To complete the proof of Eq.\ \eqref{eq:equivalence_ADM_Komar}, it is enough
to substitute Eqs.\ \eqref{eq:equality_condition} into the left-hand side of
equality \eqref{eq:improved_circularity_condition} and make use of Eqs.\ 
\eqref{eq:psi_chi_binary}.

\begin{widetext}
\section{Energy along cirular orbits
as a function of the parameter $x$ up to 10PN order} \label{app:10PN}

The coefficients $e_i$ (for $i=5,\,\ldots,\,11$) entering formula 
\eqref{eq:E_x} for energy $E$ along cirular orbits as a function of the 
parameter $x$ read
\begin{eqnarray*}
e_5 &=&
\frac{3969}{256}
+ \frac{105553}{768} \nu
- \frac{799673}{2304} \nu^2
+ \frac{502145}{3456} \nu^3
+ \frac{135247}{62208} \nu^4,
\\[2ex]
e_6 &=&
\frac{45927}{1024}
+ \frac{662619}{1024} \nu
- \frac{233737}{128} \nu^2
+ \frac{1162513}{1024} \nu^3
+ \frac{56481}{1024} \nu^4
- \frac{3777}{1024} \nu^5 ,
\\[2ex]
e_7 &=&
\frac{264627}{2048}
+ \frac{16343855}{6144} \nu
- \frac{136513223}{18432} \nu^2
+ \frac{67182599}{27648} \nu^3
+ \frac{249565613}{31104} \nu^4 
- \frac{766814851}{497664} \nu^5
- \frac{148888223 }{13436928} \nu^6,
\\[2ex]
e_8 &=&
\frac{12196899}{32768}
+ \frac{993480761}{98304} \nu
- \frac{403684801}{16384} \nu^2
- \frac{19713422131}{884736} \nu^3
+ \frac{532659868169}{3981312} \nu^4
- \frac{1096304004055}{11943936} \nu^5
\nonumber\\[2ex]&&
+ \frac{42034682027}{17915904} \nu^6
+ \frac{29197027769}{644972544} \nu^7,
\\[2ex]
e_9 &=&
\frac{70366725}{65536}
+ \frac{2394150375}{65536} \nu
- \frac{13003419935}{196608} \nu^2
- \frac{9560242585}{32768} \nu^3
+ \frac{248408863105}{196608} \nu^4
- \frac{129128080475}{98304} \nu^5
\nonumber\\[2ex]&&
+ \frac{30481497595}{98304} \nu^6
+ \frac{189766295}{16384} \nu^7
+ \frac{1599565 }{65536} \nu^8,
\\[2ex]
e_{10} &=&
\frac{813439341}{262144}
+ \frac{100457157571}{786432} \nu
- \frac{69694465361}{589824} \nu^2
- \frac{44773413415657}{21233664} \nu^3
+ \frac{558727148465215}{63700992} \nu^4
\nonumber\\[2ex]&&
- \frac{2097054471026245}{191102976} \nu^5
+ \frac{500159286898463}{143327232} \nu^6
+ \frac{858674754130903}{2579890176} \nu^7
- \frac{233795515077109}{5159780352} \nu^8
\nonumber\\[2ex]&&
- \frac{177721652120353}{417942208512} \nu^9,
\\[2ex]
e_{11} &=&
\frac{4710988269}{524288}
+ \frac{228788885795}{524288} \nu
+ \frac{411952929451}{4718592} \nu^2
- \frac{28395688669829}{2359296} \nu^3
+ \frac{3163918218071455}{63700992} \nu^4
\nonumber\\[2ex]&&
- \frac{8137959523310875}{127401984}\nu^5
+ \frac{6129421057080619}{1146617856} \nu^6
+ \frac{9328338011915891}{322486272} \nu^7
- \frac{180325112745666065}{30958682112} \nu^8
\nonumber\\[2ex]&&
- \frac{11728912802084585}{278628139008} \nu^9
+ \frac{1098645689138995}{2507653251072} \nu^{10}.
\end{eqnarray*}
\end{widetext}

\end{document}